\documentclass[12pt]{article}
\usepackage[T1]{fontenc} 
\usepackage[utf8]{inputenx}
\usepackage{babel} 
\usepackage{url}
\usepackage{geometry}
\usepackage{amsmath}
\usepackage{amsfonts}
\usepackage{amstext}
\usepackage{amssymb}
\usepackage{amsthm}
\usepackage{amscd}
\usepackage{mathrsfs}
\usepackage{xcolor}
\definecolor{myurlcolor}{rgb}{0,0,0.4}
\definecolor{mycitecolor}{rgb}{0,0.5,0}
\definecolor{myrefcolor}{rgb}{0.5,0,0}
\usepackage{graphicx}
\usepackage{tikz}
\usepackage{tikz-cd}
\usepackage{etaremune}
\usepackage{latexsym}
\usepackage{braket}
\usepackage{soul}
\setstcolor{blue}
\usepackage{comment}
\usepackage{etoolbox}
\usepackage{makeidx}
\usepackage{dsfont}
\usepackage{enumitem} 
\usepackage{caption}
\usepackage{lmodern}
\usepackage{textcomp}
\usepackage{totcount}
\usepackage{blindtext}

\usepackage[pagebackref,draft=false]{hyperref}
\hypersetup{colorlinks,linkcolor=myrefcolor,citecolor=mycitecolor,urlcolor=myurlcolor}

\usepackage[capitalize]{cleveref}

\newtheorem{remark}{Remark}

\newtheorem*{proof*}{Proof}
\newtheorem{problem}{Problem}

\newcommand{\grit}[1]{{\bfseries {\itshape {#1}}}}

\title{Can \v{C}encov meet Petz?}

\author{F. M. Ciaglia$^{1,4}$ \href{https://orcid.org/0000-0002-8987-1181}{\includegraphics[scale=0.7]{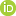}}, F. Di Cosmo$^{1,2,3}$ \href{https://orcid.org/0000-0003-0256-5913}{\includegraphics[scale=0.7]{ORCID.png}}, L. González-Bravo$^{1,5}$ \href{https://orcid.org/0000-0002-4382-7978}{\includegraphics[scale=0.5]{ORCID.png}}}

\begin{document}

\maketitle 

\noindent
{\footnotesize $^{1}$  Department of Mathematics, University Carlos III de Madrid, Legan\'es, Madrid, Spain}  \\
{\footnotesize $^{2}$ ICMAT, Instituto de Ciencias Matem\'{a}ticas (CSIC-UAM-UC3M-UCM)}

\bigskip
\noindent
{\footnotesize $^{3}$\texttt{fciaglia[at]math.uc3m.es} \quad $^{4}$\texttt{fcosmo[at]math.uc3m.es} \quad $^{5}$\texttt{lauragon[at]math.uc3m.es}}

\tableofcontents

\begin{abstract}
	We discuss how to exploit the recent formulation of classical and quantum information geometry in terms of normal states on $W^{*}$-algebras to formulate a problem that unifies \v{C}encov's theorem and Petz's theorem.
	%%%%%%%%%%%%%%%%%%%%%%%
\end{abstract}

\section{Introduction}\label{sec: introduction}

\v{C}encov's\footnote{Sometimes, the name \v{C}encov is also spelled Chentsov.} and Petz's theorems are two pillars of classical and quantum information geometry, respectively, but  their mathematical formulation is quite different as it relies on the mathematics of probability distributions in the classical case, and on the mathematics of quantum states in the quantum case.
%%%%%%%%%%%%%%%%%%%%%%%%%%%%%%
Recently, $W^{*}$-algebras and normal states on them are being employed as a common mathematical framework unifying classical and quantum information geometry \cite{C-DN-J-S-2023,C-I-J-M-2019,C-J-S-2020-02,C-J-S-2020,C-J-S-2022,Kostecki-2011,Kostecki-2016}, and in this short contribution we would like to exploit this mathematical framework in order to formulate a problem that unifies those answered by \v{C}encov and Petz, respectively.
%%%%%%%%%%%%%%%%%%%%%%%%%%%%%%%%
At this purpose, let us briefly recall \v{C}encov's and Petz's results.
%%%%%%%%%%%%%%%%%%%%%%%%%%%%%%

\v{C}encov's theorem proves that, among all possible families of Riemannian metric tensors on (the interior of) finite-dimensional simplexes, the Fisher-Rao metric tensor is the only one (up to an overall positive constant) which is invariant with respect to the class of Markov maps known as  congruent embeddings \cite{Cencov-1982}.
%%%%%%%%%%%%
Specifically, let $\mathcal{X}_{(n+1)}$ denote a discrete set with $(n+1)$-elements.
%%%%%%%%%%%%%%%%%%%
It is well-known that the space of probability measures on $\mathcal{X}_{(n+1)}$ can be identified with the $n$-simplex
\begin{equation}\label{eqn: n-simplex}
	\overline{\Delta}_{n }:=\left\{\mathbf{p}=(p^{1},\cdots , p^{n+1})\in\mathbb{R}^{n+1}\;|\quad p^{j}\geq 0,\; \sum_{j=1}^{n+1}p^{j}=1\right\}.
\end{equation}
%%%%%%%%%%%%%%%%%
The $n$-simplex is a manifold with corners and its highest-dimensional stratum $\Delta_{n}$ is composed by all those $\mathbf{p}\in\overline{\Delta}_{n}$ such that $p^{j}>0$ for $j\in[1,\cdots, n+1]$.
%%%%%%%%%%%%%%
A linear map $M\colon \mathbb{R}^{n+1}\rightarrow \mathbb{R}^{m+1}$ is called a \grit{Markov map} if $M(\overline{\Delta}_{n})\subseteq \overline{\Delta}_{m}$, and a Markov map $M$ is called faithful if $M(\Delta_{n})\subseteq \Delta_{m}$. 
%%%%%%%%%%%%%%%%%%%%%
A faithful Markov map is called a \grit{congruent embedding} if $M(\Delta_{n})\subseteq \Delta_{m}$ and it admits a left-inverse $N$ which is still a faithful Markov map.
%%%%%%%%%%%%%%%%%%
A characterization of congruent embedding can be found in \cite{Campbell-1986,Cencov-1982}.
%%%%%%%%%%%%%%%%%%%

\begin{problem}\label{prob: Cencov}
	Determine all families $\{G_{n}\}$, where $n>0$ and $G_{n}$ is a Riemannian metric tensor on $\Delta_{n}$, satisfying the invariance property
	\begin{equation}
		M^{*}G_{m} = G_{n}
	\end{equation}
	for all congruent embeddings $M$.
\end{problem}

\v{C}encov investigated problem \ref{prob: Cencov} and showed that, up to an overall positive multiplicative constant, there is only one such family determined by
\begin{equation}
	G_{n}=\sum_{j=1}^{n+1}\frac{\mathrm{d}p^{j}\otimes \mathrm{d}p^{j}}{p^{j}},
\end{equation}
where the $p^{j}$'s are the standard Cartesian coordinates on $\mathbb{R}^{n}$ associated with the standard basis and with respect to which the $n$-simplex is defined (see equation \eqref{eqn: n-simplex}) \cite{Cencov-1982,Fujiwara-2023}.
%%%%%%%%%%%%%%%%
The Riemannian metric $G_{n}$ coincides with the Fisher-Rao metric tensor on $\Delta_{n}$, ubiquitous in classical information geometry, statistics, and estimation theory \cite{A-J-L-S-2017}.
%%%%%%%%%%%%%%

After tackling the classical case, \v{C}encov and Morozova started an investigation of the quantum case \cite{C-M-1987,C-M-1991}.
%%%%%%%%%%%%%%%
They provided a preliminary partial solution of the problem that was later fully developed by Petz in the finite-dimensional case.
%%%%%%%%%%%%%%%%%%%%%%%%%%%%%%
Petz's theorem is a sort of quantum counterpart of \v{C}encov's result for it classifies, among all possible families of Riemannian metric tensors on the manifolds of faithful quantum states in finite dimensions, the ones that are monotone with respect to the completely-positive and trace-preserving (CPTP) maps \cite{Petz-1996}.
%%%%%%%%%%%%%
Specifically, let us consider a quantum system with a finite-dimensional Hilbert space $\mathcal{H}\cong\mathbb{C}^{n}$. 
%%%%%%%%%%%%%%%%
A quantum state $\rho$ is then a density operator on $\mathcal{H}$, namely, a self-adjoint operator in the algebra $\mathcal{B}(\mathcal{H})$ of bounded linear operators which has unit trace and is positive semi-definite, meaning that $\langle\psi\mid \rho\mid\psi\rangle\geq 0$ for every $\psi\in\mathcal{H}$.
%%%%%%%%%%%%%%%%%
A quantum state is called \grit{faithful} if $\langle\psi\mid \rho\mid\psi\rangle = 0$ implies that $\psi$ is the zero vector in $\mathcal{H}$.
%%%%%%%%%%%%%%%%%%%%
Note that, in finite dimensions, a faithful quantum state is invertible as a linear operator (the same is no-longer true in infinite dimensions because density operators are required to be trace-class so that they never admit a bounded inverse).
%%%%%%%%%%%%%%%%%%%%%
The space of faithful quantum states is denoted by $\mathscr{S}_{f}(\mathcal{H})$, and it is a smooth manifold which is also an homogeneous space for the Lie group $GL(\mathcal{H})$ \cite{C-C-I-M-V-2019,C-DC-I-L-M-2017,G-K-M-2005}.
%%%%%%%%%%%%%%%%%%%%
A linear map $\Phi\colon \mathcal{B}(\mathcal{H})\rightarrow\mathcal{B}(\mathcal{K})$ is called positive if it sends positive operators to positive operators, and it is called completely-positive (CP) if the map $\Phi\otimes\mathrm{1}_{n}\colon \mathcal{B}(\mathcal{H})\otimes \mathrm{M}_{n}(\mathbb{C})\rightarrow \mathcal{B}(\mathcal{K})\otimes \mathrm{M}_{n}(\mathbb{C})$, where $\mathrm{1}_{n}$ is the identity map on the space of square complex matrices $\mathrm{M}_{n}(\mathbb{C})=\mathcal{B}(\mathbb{C}^{n})$, is positive for all $n$ \cite{Choi-1974}.
%%%%%%%%%%%%%%%%%%%%
A CP maps $\Phi$ is called trace-preserving (TP) if $\mathrm{Tr}_{\mathcal{K}}(\Phi(\mathbf{x}))=\mathrm{Tr}_{\mathcal{H}}(\mathbf{x})$ for all $\mathbf{x}\in \mathcal{B}(\mathcal{H})$.
%%%%%%%%%%%%%%%%%%%%%%
Of course, a completely-positive and trace-preserving (CPTP) map $\Phi$ sends quantum states into quantum states, and if $\Phi$ sends faithful states into faithful states, meaning that $\Phi(\mathscr{S}_{f}(\mathcal{H}))\subseteq\mathscr{S}_{f}(\mathcal{K})$, then it is called \grit{faithful} and referred to as a fCPTP map.
%%%%%%%%%%%%%%%%%%%%%%

It is worth noting that faithful positive maps are also called \textit{strictly positive} \cite{Bhatia-2007}, and have been recently shown to possess a sort of ``universal kernel'' \cite{vomEnde-2020}.

\begin{problem}\label{prob: petz}
	Determine all families $\{G_{\mathcal{H}}\}$, where $\mathcal{H}$ is a finite-dimensional Hilbert space and $G_{\mathcal{H}}$ is a Riemannian metric tensor on $\mathscr{S}_{f}(\mathcal{H})$, satisfying the monotonicity property
	\begin{equation}\label{eqn: monotonicity petz}
		\Phi^{*}G_{\mathcal{K}}\leq G_{\mathcal{H}}
	\end{equation}
	for all fCPTP maps.
\end{problem}

\begin{remark}
	A  quantum congruent embedding would be an fCPTP map $\Phi$ possessing a left-inverse $\Psi$ which is itself an fCPTP map.
	%%%%%%%%%%%%%%%%%
	The most general form of such a map is $\Phi(\rho)=\mathbf{U}(\rho\otimes\sigma)\mathbf{U}^{\dagger}$, where $\sigma$ is a fixed faithful quantum state on $\mathcal{K}$ and $\mathbf{U}$ a unitary operator on $\mathcal{H}\otimes \mathcal{K}$ \cite{N-S-2007}.
	%%%%%%%%%%%%%%%%
	A double application  of equation \eqref{eqn: monotonicity petz} forces every monotone family to be invariant with respect to quantum congruent embeddings.
	%%%%%%%%%%%%%%%%
\end{remark}

\v{C}encov and Morozova gave a preliminary definition and partial solution of  problem \ref{prob: petz} \cite{C-M-1987,C-M-1991}.
%%%%%%%%%%%%%%%%%%%%
Petz investigated problem \ref{prob: petz} and showed that every admissible monotone family is of the type $\{G_{\mathcal{H}}^{f}\}$ where $f$ is an operator monotone function $f\colon[0,\infty)\rightarrow[0,\infty)$ satisfying $f(t)=t f(t^{-1})$ \cite{Petz-1996}.
%%%%%%%%%%%%%%%%%%%%%%%%%%
We refer to this family as the Morozova-\v{C}encov-Petz family of monotone metric tensors.
%%%%%%%%%%%%%%%%%%%%%%%%
In particular, the explicit expression of the Riemannian metric tensor $G_{\mathcal{H}}^{f}$ reads
\begin{equation}\label{eqn: petz metric}
	(G_{\mathcal{H}}^{f})_{\rho}(\mathbf{a},\mathbf{b})=\mathrm{Tr}_{\mathcal{H}}\left(\mathbf{a}(K^{f}_{\rho})^{-1}(\mathbf{b})\right),
\end{equation}
where $\mathbf{a},\mathbf{b}\in T_{\rho}\mathscr{S}_{f}(\mathcal{H})\subset \mathcal{B}(\mathcal{H})$ are trace-less, and the superoperator $K_{\rho}^{f}$ is given by
\begin{equation}
	K_{\rho}^{f}=f\left(L_{\rho}R_{\rho}^{-1}\right)\,R_{\rho},
\end{equation}
with $L_{\rho}(\mathbf{b})=\rho\mathbf{b}$ and $R_{\rho}(\mathbf{b})=\mathbf{b}\rho$.
%%%%%%%%%%%%%%%%%%%%%

In the rest of this work we exploit the formalism of $W^{*}$-algebras in order to formulate a problem that encompasses and unifies problem \ref{prob: Cencov} and problem \ref{prob: petz}, and discuss a conjecture concerning a possible solution of said problem.
%%%%%%%%%%%%%%%%%%%%%%%%%%

\section{Information geometry and $W^{*}$-algebras}

Even if the mathematical frameworks underlying \v{C}encov's and Petz's theorems may seem quite different at first sight, they can be unified in the context of $W^{*}$-algebras, namely, $C^{*}$-algebras admitting a pre-dual \cite{takesaki-2002}.
%%%%%%%%%%%%%%%%%%%
Roughly speaking, a $C^{*}$-algebra is a functional-analytic abstraction of the algebra $\mathrm{M}_{n}(\mathbb{C})$ of square complex matrices.
%%%%%%%
Specifically, a $C^{*}$-algebra is a quintuple $\mathscr{A}\equiv (A,+,\cdot,\lVert\cdot \rVert,\dagger)$ where $(A,+,\cdot,\lVert\cdot \rVert)$ is a complex, associative Banach algebra endowed with a map $\dagger\colon A\rightarrow A$, often written as $\dagger(\mathbf{x})\equiv \mathbf{x}^{\dagger}$, which is norm-continuous and satisfies the following properties: 
\begin{align}
	(\alpha \mathbf{x} + \beta \mathbf{y})^{\dagger} & = \overline{\alpha}\mathbf{x}^{\dagger} + \overline{\beta}\mathbf{y}^{\dagger} \quad	\forall \mathbf{x},\mathbf{y}\in A,\;\;\forall \alpha,\beta\in\mathbb{C}\\ 
	(\mathbf{xy})^{\dagger}&=\mathbf{y}^{\dagger}\,\mathbf{x}^{\dagger} \quad	\forall \mathbf{x},\mathbf{y}\in A\\
	(\mathbf{x}^{\dagger})^{\dagger} &= \mathbf{x}\quad	\forall \mathbf{x}\in A\\
	\lVert \mathbf{x}\mathbf{x}^{\dagger}\rVert &=  \lVert\mathbf{x}\rVert^{2} .
\end{align}
%%%%%%%%%%%%%%
A prototypical examples of unital, Abelian $C^{*}$-algebra is the algebra $C(\mathcal{X})$ of complex-valued, continuous function on the topological space $\mathcal{X}$, where the norm is the sup-norm and $\dagger$ is just complex conjugation, while the prototypical example of unital, non-Abelian $C^{*}$-algebra is the algebra $\mathcal{B}(\mathcal{H})$ of bounded linear operators on the complex Hilbert space $\mathcal{H}$, where the norm is the operator norm and $\dagger$ is the operator adjoint (in particular, when $\mathcal{H}\cong \mathbb{C}^{n}$ then $\mathcal{B}(\mathcal{H})\cong\mathrm{M}_{n}(\mathbb{C})$).
%%%%%%%%%%%%%%%%%

An element $\mathbf{x}\in\mathscr{A}$ is called self-adjoint if $\mathbf{x}^{\dagger}=\mathbf{x}$, while it is called positive if $\mathbf{x}=\mathbf{y}\mathbf{y}^{\dagger}$ for some $\mathbf{y}\in \mathscr{A}$. 
%%%%%%%%%%
Note that positive elements are necessarily self-adjoint.
%%%%%%%%%%%%
Positive elements are positive functions when $\mathscr{A}=C(\mathcal{X})$, and are positive semidefinite operators when $\mathscr{A}=\mathcal{B}(\mathcal{H})$.
%%%%%%%%%%%%%

A $W^{*}$-algebra $\mathscr{A}$ is a $C^{*}$-algebra which is itself the dual space of a Banach space called the \grit{pre-dual} of $\mathscr{A}$ and denoted by $\mathscr{A}_{*}$.
%%%%%%%%%%%%%%%%%%%%%%
The notion of $W^{*}$-algebra is an algebraic abstraction of that of a von Neumann algebra, namely, an involutive subalgebra of some $\mathcal{B}(\mathcal{H})$  which is equal to its double commutant (or closed in the weak operator topology), in the sense that every von Neumann algebra is a $W^{*}$-algebra and every $W^{*}$-algebra can be concretely represented as a von Neumann algebra \cite{Sakai-1956}.
%%%%%%%%%%%
Consequently,  every $W^{*}$-algebras $\mathscr{A}$ is unital because, when represented as a von Neumann algebra, it must be equal to its double commutant which always contains the identity operator. 
%%%%%%%%%%%%%%%%%%%%%%%%
Of course, every finite-dimensional $C^{*}$-algebra is a $W^{*}$-algebra, while the same is not true in infinite dimensions.
%%%%%%%%%%%%%%%%%%
For instance, the $C^{*}$-algebra $\mathcal{K}(\mathcal{H})$ of compact linear operators on the complex Hilbert space $\mathcal{H}$ is a $C^{*}$-algebra which is not a $W^{*}$-algebra (for instance, it does not contain the identity operator).
%%%%%%%%%%%%%%%
The prototypical example of an Abelian $W^{*}$-algebra is $\mathcal{L}^{\infty}(\mathcal{X},\mu)$, that is, the algebra of (equivalence classes of) complex-valued measurable functions on the measure space $(\mathcal{X},\mu)$ which are $\mu$-essentially bounded.
%%%%%%%%%%%%%%%
From standard results in functional analysis, the pre-dual of  $\mathcal{L}^{\infty}(\mathcal{X},\mu)$ can be identified with  $\mathcal{L}^{1}(\mathcal{X},\mu)$ according to
\begin{equation}
	\langle f\mid \xi \rangle :=\int_{\mathcal{X}}f(x)\,\xi(x)\,\mathrm{d}\mu(x),
\end{equation}
where $f\in \mathcal{L}^{\infty}(\mathcal{X},\mu)$ and $\xi\in\mathcal{L}^{1}(\mathcal{X},\mu)$.
%%%%%%%%%%%%%%%%%%%%%%%%%%%
The prototypical example of non-Abelian $W^{*}$-algebra is again $\mathcal{B}(\mathcal{H})$, which is the dual space of the space $\mathcal{TC}(\mathcal{H})$ of trace-class linear operators on $\mathcal{H}$ (therefore, $\mathcal{B}(\mathcal{H})$ is also the double dual of the $C^{*}$-algebra $\mathcal{K}(\mathcal{H})$ of compact linear operators on $\mathcal{H}$).
%%%%%%%%%%%%%%%%%%%%%%%%%%%
In this case, the dual pairing is given by the trace on $\mathcal{H}$ according to 
\begin{equation}
	\langle \mathbf{x}\mid \rho\rangle := \mathrm{Tr}_{\mathcal{H}}(\mathbf{x}\,\rho),
\end{equation}
where $\mathbf{x}\in\mathcal{B}(\mathcal{H})$ and $\rho\in\mathcal{TC}(\mathcal{H})$.
%%%%%%%%%%%%%%%%%%%%%%%%%

The way in which $W^{*}$-algebras provide a unifying framework for classical and quantum information geometry is through the notion of normal states.
%%%%%%%%%%%%%
Given a $C^{*}$-algebra $\mathscr{A}$, a positive linear functional $\rho$ is a linear functional in the dual space $\mathscr{A}^{*}$ such that 
\begin{equation}
	\rho(\mathbf{x}^{\dagger}\mathbf{x})\geq 0 \quad \forall \mathbf{x}\in\mathscr{A},
\end{equation}
which means that $\rho$ takes non-negative values on positive elements.
%%%%%%%%%%%%%%%%%%%%%%%%%
A positive linear functional is called \grit{faithful} if $\rho(\mathbf{x}^{\dagger}\mathbf{x})=0$ implies $\mathbf{x}=\mathbf{0}$.
%%%%%%%%%%%%%%%%%%%%%%%%%%%%%
The space of positive linear functionals on $\mathscr{A}$ is denoted by $\mathscr{P}(\mathscr{A})$, while the space of faithful positive linear functionals is denoted by $\mathscr{P}_{f}(\mathscr{A})$.
%%%%%%%%%%%%%%%%%%%%%%%%%%%%%%
A positive linear functional $\rho$ is a \grit{state} if it is normalized in the sense that $\lVert\rho\rVert=1$.
%%%%%%%%%%%%%%%%%%
In the particular case in which $\mathscr{A}$ admits a unit $\mathbb{I}$, the norm of a positive linear functional can be proved to satisfy the equality $\lVert\rho  \rVert=\rho(\mathbb{I})$ \cite{takesaki-2002}.
%%%%%%%%%%%%%%%%%%%%%%
A state is called \grit{faithful} if it is faithful as a positive linear functional.
%%%%%%%%%%%%%%%%%%%%%
The space of states on $\mathscr{A}$ is denoted by $\mathscr{S}(\mathscr{A})$, while the space of faithful states is denoted by $\mathscr{S}_{f}(\mathscr{A})$.
%%%%%%%%%%%%%%%%%%%%%

Given a state $\rho\in\mathscr{S}(\mathscr{A})$, we can always build a Hilbert space $\mathcal{H}_{\rho}$ which is naturally associated with $\rho$ by means of the so-called Gelfand-Naimark-Segal (GNS) construction \cite{B-R-1987-1}.
%%%%%%%%%%%%%%%%%%%%
Specifically, we introduce a pre-Hilbert space structure on $\mathscr{A}$ setting
\begin{equation}\label{eqn: GNS scalar product}
	\langle\mathbf{x}\mid \mathbf{y}\rangle_{\rho}:=\rho(\mathbf{x}^{\dagger}\mathbf{y}),
\end{equation}
and then introduce the subset $N_{\rho}=\{\mathbf{x}\in\mathscr{A}\;|\;\;\rho(\mathbf{x}^{\dagger}\mathbf{x})=0\}$, which is a left ideal in $\mathscr{A}$ called the Gel'fand ideal.
%%%%%%%%%%%%%%%%%%%%%
The scalar product $\langle\cdot \mid \cdot\rangle_{\rho}$ descends to the quotient $\mathscr{A}/N_{\rho}$ on which it is non-degenerate by construction.
%%%%%%%%%%%%%%%%%%%%%%
Finally, we obtain the (GNS) Hilbert space $\mathcal{H}_{\rho}$ as the Hilbert-space-completion of $\mathscr{A}/N_{\rho}$.
%%%%%%%%%%%%%%%%%%%%%%

If $\mathscr{A}$ is a $W^{*}$-algebra, a linear functional $\hat{\xi}\in \mathscr{A}^{*}$ is called \grit{normal} if it is the image $\hat{\xi}=i(\xi)$ of an element $\xi \in \mathscr{A}_{*}$ through the canonical immersion $i$ of the pre-dual $\mathscr{A}_{*}$ into its double dual $\mathscr{A}^{*}$.
%%%%%%%%%%%%%%%%
In the following, we will often identify $\mathscr{A}_{*}$ with its image in $\mathscr{A}^{*}$ through $i$ for the sake of notational simplicity.
%%%%%%%%%%%%%%%%%%%%%%
The space of normal (faithful) positive linear functionals on $\mathscr{A}$ is denoted by $\mathscr{P}_{n}(\mathscr{A})$ ($\mathscr{P}_{nf}(\mathscr{A})$), while the space of normal (faithful) states on $\mathscr{A}$ is denoted by $\mathscr{S}_{n}(\mathscr{A})$ ($\mathscr{S}_{nf}(\mathscr{A})$).
%%%%%%%%%%%%%%%%%
Of course, in the finite-dimensional case, all linear functionals are normal.
%%%%%%%%%%%%%%%%

A linear map $\Psi\colon \mathscr{A}\rightarrow \mathscr{B}$ between $C^{*}$-algebras is called \grit{normal} if its dual map $\Psi^{*}$ is such that $\Psi^{*}(\mathscr{B}_{*})\subseteq \mathscr{A}_{*}$; it is called positive if it sends positive elements into positive elements,and it is called completely-positive (CP) if the map $\Psi\otimes\mathrm{1}_{n}\colon \mathscr{A}\otimes \mathrm{M}_{n}(\mathbb{C})\rightarrow \mathscr{B}\otimes \mathrm{M}_{n}(\mathbb{C})$, where $\mathrm{1}_{n}$ is the identity map on the $W^{*}$-algebra of square complex matrices, is positive for all $n$ \cite{Choi-1974}.
%%%%%%%%%%%%%%%%%%%%%%%%%%%%%
When either $\mathscr{A}$ or $\mathscr{B}$ is Abelian, all positive maps are automatically completely-positive.
%%%%%%%%%%%%%%%%%%%%%%%%%%%%
The positive map $\Psi$ is called unital if $\Psi(\mathbb{I}_{\mathscr{A}})=\mathbb{I}_{\mathscr{B}}$, and a completely-positive map which is unital will be denoted called a CPU map.
%%%%%%%%%%%%%%%%%%%%%%%
The dual map $\Psi^{*}$ of the normal CPU map $\Psi$ preserves normal states in the sense that $\Psi^{*}(\mathscr{S}_{n}(\mathscr{B}))\subseteq \mathscr{S}_{n}(\mathscr{A})$, and $\Psi$ is in addition called faithful if it preserves faithful states.
%%%%%%%%%%%%%%%%%%%%%%%%
Maps that are normal, CPU and faithful will be called fnCPU maps.
%%%%%%%%%%%%%%%%%%%%%%%%%%%%%%

\section{Can \v{C}encov meet Petz?}

We are now in a position to present a unified mathematical framework for classical and quantum information geometry in which \v{C}encov's and Petz's theorems would appear as particular cases of a yet to be proved general theorem.
%%%%%%%%%%%%%%%%%%%%%%%%%%%%%%%
The idea is to consider finite-dimensional $W^{*}$-algebras and the manifolds of normal faithful states as the unifying objects for probability distributions and quantum states.
%%%%%%%%%%%%%%%%
Indeed, the connection with classical information geometry follows from the fact that the space of normal states of $\mathscr{A}=\mathcal{L}^{\infty}(\mathcal{X},\mu)$ is identified with the space of probability measures on $(\mathcal{X},\mu)$ which are absolutely continuous with respect to $\mu$.
%%%%%%%%%%%%%%%%%%%%
Consequently, when considering the finite discrete space $\mathcal{X}_{n}$ endowed with the counting measure $\mu$, the space $\mathscr{S}_{nf}(\mathscr{A})$ of faithful normal states on $\mathscr{A}=\mathcal{L}^{\infty}(\mathcal{X}_{n},\mu)$ coincides with the interior $\Delta_{n}$ of the $n$-simplex appearing as the geometrical background of \v{C}encov's theorem discussed in section \ref{sec: introduction}.
%%%%%%%%%%%%%%%%%%%%%
Analogously, the connection with quantum information geometry follows from the fact that the space of normal states of $\mathscr{A}=\mathcal{B}(\mathcal{H})$ is identified with the space of density operators on $\mathcal{H}$, that is, with the space of quantum states of quantum mechanics.
%%%%%%%%%%%%%%%%%%%%%
Therefore, when $\mathcal{H}\cong \mathbb{C}^{n}$, the space $\mathscr{S}_{nf}(\mathscr{A})$ of faithful normal states on $\mathscr{A}=\mathcal{B}(\mathcal{H})$ coincides with the space of faithful (invertible) quantum states appearing as the geometrical background of Petz's theorem discussed in section \ref{sec: introduction}.
%%%%%%%%%%%%%%%%%%%
Then, instead of faithful Markov maps and fCPTP maps, we consider the class of maps which are dual to nfCPU maps.
%%%%%%%%%%%%%%%%%%%%%
When $\mathscr{A}\cong\mathcal{L}^{\infty}(\mathcal{X}_{n},\mu)$ then these maps reduce to the faithful Markov maps appearing in \v{C}encov's theorem, while when $\mathscr{A}\cong \mathcal{B}(\mathcal{H})$, with $\mathcal{H}\cong\mathbb{C}^{n}$, these maps reduce to the fCPTP maps appearing in Petz's theorem.
%%%%%%%%%%%%%%%%%%%%%%%%%%%%%%%%
Finally, we may formulate the following problem that unifies the ones discussed by \v{C}encov and Petz in a single framework:

\begin{problem}\label{pro: Cencov+petz}
	Determine all families $\{G_{\mathscr{A}}\}$, where $\mathscr{A}$ is a finite-dimensional $W^{*}$-algebra and $G_{\mathscr{A}}$ is a Riemannian metric tensor on $\mathscr{S}_{nf}(\mathscr{A})$, satisfying the monotonicity property
	\begin{equation}\label{eqn: monotonicity property W*}
		\Phi^{*}G_{\mathscr{A}}\leq G_{\mathscr{B}}
	\end{equation}
	for every map $\Phi$ which is the dual of a nfCPU map $\Psi\colon \mathscr{A}\rightarrow \mathscr{B}$.
\end{problem}

\begin{remark}
	If $\Phi$ admits a left-inverse which is still the dual of a nfCPU, then equation \eqref{eqn: monotonicity property W*} becames an equality.
\end{remark}

At the moment, we do not have a complete solution to problem \ref{pro: Cencov+petz}, but what we can show is that Petz's classification, once suitably reformulated in the $W^{*}$-algebraic language, provides us with families of Riemannian metric tensors satisfying the monotonicity property in equation \eqref{eqn: monotonicity property W*}.
%%%%%%%%%%%%%%%%%%%%%%%%%%%%%
At this purpose, we note that, being $\rho$ invertible as a density operator, both the left and right multiplication operators $L_{\rho}$ and $R_{\rho}$ are invertible as superoperators acting on $\mathcal{B}(\mathcal{H})$.
%%%%%%%%%%%%
Therefore, every $\mathbf{a}\in T_{\rho}\mathscr{S}_{f}(\mathcal{H})$ can be written as
\begin{equation}
	\mathbf{a}=\{\rho,\mathbf{v}_{\rho}\}\equiv\frac{1}{2}\left(\rho\,\mathbf{v}_{\rho} + \mathbf{v}_{\rho}\,\rho\right)= \frac{R_{\rho}}{2}(L_{\rho}R_{\rho}^{-1} + \mathbf{1}_{\mathcal{B}(\mathcal{H})})(\mathbf{v}_{\rho}),
\end{equation}
and $\mathrm{Tr}_{\mathcal{H}}(\rho\,\mathbf{v}_{\rho})=0$.
%%%%%%%%%%%%%%
Then, we recall that the superoperator $L_{\rho}R_{\rho}^{-1}$ appears in the framework of $W^{*}$-algebras as the generator $\Delta_{\rho}$ of the modular automorphism associated with the state $\rho$ on the GNS Hilbert space associated with $\rho$ \cite{Petz-1986}.
%%%%%%%%%%%%%%%%%%
Therefore, we can rewrite equation \eqref{eqn: petz metric} as
\begin{equation}
	(G_{\mathcal{H}}^{f})_{\rho}(\mathbf{a},\mathbf{b})=\mathrm{Tr}_{\mathcal{H}}\left(\mathbf{a}(K^{f}_{\rho})^{-1}(\mathbf{b})\right)=\frac{1}{2}\mathrm{Tr}_{\mathcal{H}}\left(\rho\left\{\mathbf{v}_{\rho},\,F(\Delta_{\rho})(\mathbf{w}_{\rho})\right\}\right)
\end{equation}
where
\begin{equation}
	F(\Delta_{\rho}):=(f(\Delta_{\rho}))^{-1}(L_{\rho}R_{\rho}^{-1} + \mathbf{1}_{\mathcal{B}(\mathcal{H})}),
\end{equation}
so that, recalling the GNS scalar product in equation \eqref{eqn: GNS scalar product}, we conclude that
\begin{equation}\label{eqn: petz GNS}
	(G_{\mathcal{H}}^{f})_{\rho}(\mathbf{v}_{\rho},\mathbf{w}_{\rho})= \Re\left(\langle\mathbf{v}_{\rho}\mid F(\Delta_{\rho})(\mathbf{w}_{\rho})\rangle_{\rho}\right).
\end{equation}
%%%%%%%%%%%%%%
The right-hand-side of equation \eqref{eqn: petz GNS} is well-defined for every $W^{*}$-algebra so that we can define the family $\{G_{\mathscr{A}}^{F}\}$ of Riemannian metric tensors setting
\begin{equation}\label{eqn: monotone metrics W*algebras}
	(G_{\mathscr{A}}^{F})_{\rho}(\mathbf{v}_{\rho},\mathbf{w}_{\rho})= \Re\left(\langle\mathbf{v}_{\rho}\mid F(\Delta_{\rho})(\mathbf{w}_{\rho})\rangle_{\rho}\right)
\end{equation}
where $\mathbf{v}_{\rho},\mathbf{w}_{\rho}\in T_{\rho}\mathscr{S}_{nf}(\mathscr{A})$ and  $\langle\cdot\mid\cdot\rangle_{\rho}$ is the GNS Hilbert space product associated with $\rho$.
%%%%%%%%%%%%%
By suitably adapting the proof of theorem 4 in \cite{Petz-1985}, it is possible to prove that the elements in the family $\{G_{\mathscr{A}}^{F}\}$ satisfy the monotonicity property in equation \eqref{eqn: monotonicity property W*}, and thus are admissible monotone families (we omit the proof here because of space constraints and we refer to a forthcoming work for a detailed proof).
%%%%%%%%%%%%%%%%

\begin{remark}
	After completing this work, it has been pointed out to us that equation \eqref{eqn: monotone metrics W*algebras} essentially coincides with the real part of equation [146] in \cite{Kostecki-2016}, where, however, the author introduces a family of Morozova-\v{C}encov-Petz-like inner products on a possibly infinite-dimensional $W^{*}$-algebra. 
	%%%%%%%%%%%%%%%%%%%%%
	A deeper investigation of the connections between our work an \cite{Kostecki-2016} will be the object of future works.
	%%%%%%%%%%%%%%%
\end{remark}

\section{Conclusions}

The formalism of $W^{*}$-algebras and normal states on them provides a natural framework for discussing \v{C}encov's and Petz's problems simultaneously.
%%%%%%%%%%%%%%%%%%
In particular, solving problem \ref{pro: Cencov+petz} would allow for a simultaneous generalization of both problem \ref{prob: Cencov} and problem \ref{prob: petz} in such a way that the solutions of both these problems, as provided by \v{C}encov and Petz respectively, appear as particular cases.
%%%%%%%%%%%%%%%%%%
We conjecture that the expression appearing in equation \eqref{eqn: monotone metrics W*algebras} provides a solution for problem \ref{pro: cencov+petz}.
%%%%%%%%%%%%%%%%%%%%
One of the reasons behind our conjecture is that the appearance of a function of $\Delta_{\rho}$ as the unique departure from the GNS Hilbert product is compatible with \v{C}encov's uniqueness result in the classical case because $\Delta_{\rho}$ reduces to the identity for every faithful normal state on a commutative $W^{*}$-algebra.
%%%%%%%%%%%%%%%%%%%%
We are actively investigating the validity of this conjecture in a setting that extends the one presented here because it deals with not-necessarily finite-dimensional $W^{*}$-algebras, thus bringing to the table a whole lot of functional-analytic technicalities.
%%%%%%%%%%%%%%%%%%%%%%

%\addcontentsline{toc}{section}{References}


\begin{thebibliography}{10}
	
	\bibitem{A-J-L-S-2017}
	N.~Ay, J.~Jost, H.~V. Le, and L.~Schwachh\"{o}fer.
	\newblock {\em {Information Geometry}}.
	\newblock Springer International Publishing, 2017.
	\newblock {\href{https://doi.org/10.1007/978-3-319-56478-4}{DOI:
			10.1007/978-3-319-56478-4}}.
	
	\bibitem{Bhatia-2007}
	R.~Bhatia.
	\newblock {\em {Positive Definite Matrices}}.
	\newblock Princeton University Press, 2007.
	
	\bibitem{B-R-1987-1}
	O.~Bratteli and D.~W. Robinson.
	\newblock {\em {Operator Algebras and Quantum Statistical Mechanics I}}.
	\newblock Springer-Verlag, Berlin, second edition, 1987.
	\newblock {\href{https://doi.org/10.1007/978-3-662-03444-6}{DOI:
			10.1007/978-3-662-03444-6}}.
	
	\bibitem{Campbell-1986}
	L.~L. Campbell.
	\newblock An extended \v{C}encov characterization of the information metric.
	\newblock {\em Proceedings of the American Mathematical Society}, 98(1):135 --
	141, 1986.
	
	\bibitem{Choi-1974}
	M.~Choi.
	\newblock {A schwarz inequality for positive linear maps on
		$C^{\ast}$-algebras}.
	\newblock {\em Illinois Journal of Mathematics}, 4(3):565--574, 1974.
	\newblock {\href{https://doi.org/10.1215/ijm/1256051007}{DOI:
			10.1215/ijm/1256051007}}.
	
	\bibitem{C-C-I-M-V-2019}
	D.~Chru\'{s}ci\'{n}ski, F.~M. Ciaglia, A.~Ibort, G.~Marmo, and F.~Ventriglia.
	\newblock {Stratified manifold of quantum states, actions of the complex
		special linear group}.
	\newblock {\em Annals of Physics}, 400:221--245, 2019.
	\newblock {\href{https://doi.org/10.1016/j.aop.2018.11.015}{DOI:
			10.1016/j.aop.2018.11.015},
		\href{https://arxiv.org/abs/1811.07406}{arXiv:1811.07406 [quant-ph]}}.
	
	\bibitem{C-DC-I-L-M-2017}
	F.~M. Ciaglia, F.~Di~Cosmo, A.~Ibort, M.~Laudato, and G.~Marmo.
	\newblock {Dynamical vector fields on the manifold of quantum states}.
	\newblock {\em Open Systems \& Information dynamics}, 24(3):1740003--38, 2017.
	\newblock {\href{https://doi.org/10.1142/S1230161217400030}{DOI:
			10.1142/S1230161217400030}, \href{https://arxiv.org/abs/1707.00293}{arXiv:
			1707.00293 [quant-ph]}}.
	
	\bibitem{C-DN-J-S-2023}
	F.~M. Ciaglia, F.~Di~Nocera, J.~Jost, and L.~Schwachh\"{o}fer.
	\newblock {Parametric models and information geometry on $W^{*}$-algebras}.
	\newblock {\em Information Geometry}, 5(1), 2023.
	\newblock {\href{https://doi.org/10.1007/s41884-022-00094-6}{DOI:
			10.1007/s41884-022-00094-6}\href{https://arxiv.org/abs/2207.09396},
		{arXiv:2207.09396 [math-ph]}}.
	
	\bibitem{C-I-J-M-2019}
	F.~M. Ciaglia, A.~Ibort, J.~Jost, and G.~Marmo.
	\newblock {Manifolds of classical probability distributions and quantum density
		operators in infinite dimensions}.
	\newblock {\em Information Geometry}, 2(2):231--271, 2019.
	\newblock {\href{https://doi.org/10.1007/s41884-019-00022-1}{DOI:
			10.1007/s41884-019-00022-1}}.
	
	\bibitem{C-J-S-2020-02}
	F.~M. Ciaglia, J.~Jost, and L.~Schwachh\"{o}fer.
	\newblock {Differential geometric aspects of parametric estimation theory for
		states on finite-dimensional C*-algebras}.
	\newblock {\em Entropy}, 22(11):1332, 2020.
	\newblock {\href{https://doi.org/10.3390/e22111332}{DOI: 10.3390/e22111332},
		\href{https://arxiv.org/abs/2010.14394}{arXiv: 2010.14394 [math-ph]}}.
	
	\bibitem{C-J-S-2020}
	F.~M. Ciaglia, J.~Jost, and L.~Schwachh\"{o}fer.
	\newblock {From the Jordan product to Riemannian geometries on classical and
		quantum states}.
	\newblock {\em Entropy}, 22(06):637--27, 2020.
	\newblock {\href{https://doi.org/10.3390/e22060637}{DOI: 10.3390/e22060637},
		\href{https://arxiv.org/abs/2005.02023}{arXiv:2005.02023 [math-ph]}}.
	
	\bibitem{C-J-S-2022}
	F.~M. Ciaglia, J.~Jost, and L.~Schwachh\"{o}fer.
	\newblock {Information geometry, Jordan algebras, and a coadjoint orbit-like
		construction}.
	\newblock {\em arXiv}, 2021.
	\newblock {\href{https://arxiv.org/abs/2112.09781}{arXiv:2112.09781
			[math.DG]}}.
	
	\bibitem{Fujiwara-2023}
	A.~Fujiwara.
	\newblock {Hommage to Chentsov's theorem}.
	\newblock {\em Information Geometry}, 2023.
	\newblock {\href{https://doi.org/10.1007/s41884-022-00077-7}{DOI:
			10.1007/s41884-022-00077-7}}.
	
	\bibitem{G-K-M-2005}
	J.~Grabowski, M.~Ku{\'s}, and G.~Marmo.
	\newblock {Geometry of quantum systems: density states and entanglement}.
	\newblock {\em Journal of Physics A: Mathematical and General},
	38(47):10217--10244, 2005.
	\newblock {\href{https://doi.org/10.1088/0305-4470/38/47/011}{DOI:
			10.1088/0305-4470/38/47/011}}.
	
	\bibitem{Kostecki-2011}
	R.~P. Kostecki.
	\newblock {Quantum theory as inductive inference}.
	\newblock {\em AIP Conference Proceedings}, 1305(33), 2011.
	\newblock {\href{https://aip.scitation.org/doi/10.1063/1.3573636}{DOI:
			10.1063/1.3573636}, \href{https://arxiv.org/abs/1009.2423}{arXiv:1009.2423
			[math-ph]}}.
	
	\bibitem{Kostecki-2016}
	R.~P. Kostecki.
	\newblock {Local quantum information dynamics}.
	\newblock 2016.
	\newblock {\href{https://arxiv.org/abs/1605.02063}{arXiv:1605.02063
			[quant-ph]}}.
	
	\bibitem{C-M-1987}
	E.~A. Morozova and N.~N. \v{C}encov.
	\newblock {Markov maps in noncommutative probability theory and mathematical
		statistics}.
	\newblock In Y.~V. Prokhorov, V.~A. Statulevi\v{c}ius, V.~V. Sazonov, and
	B.~Grigelionis, editors, {\em Probability theory and mathematical statistics:
		proceedings of the Fourth Vilnius Conference}, volume~2, pages 287--310. VNU
	Science Press, Utrecht, 1987.
	\newblock {\href{https://doi.org/10.1515/9783112313985-026}{DOI:
			10.1515/9783112313985-026}}.
	
	\bibitem{C-M-1991}
	E.~A. Morozova and N.~N. \v{C}encov.
	\newblock {Markov invariant geometry on manifolds of states}.
	\newblock {\em Journal of Soviet Mathematics}, 56(5):2648--2669, 1991.
	\newblock {\href{https://doi.org/10.1007/BF01095975}{DOI: 10.1007/BF01095975}}.
	
	\bibitem{N-S-2007}
	A.~Nayak and P.~Sen.
	\newblock {Invertible quantum operations and perfect encryption of quantum
		states}.
	\newblock {\em Quantum Information \& Computation}, 07(01):103--110, 2007.
	\newblock {\href{https://doi.org/10.5555/2011706.2011712}{DOI:
			10.5555/2011706.2011712}}.
	
	\bibitem{Petz-1985}
	D.~Petz.
	\newblock {Quasi-entropies for States of a von Neumann Algebra}.
	\newblock {\em Publications of the RIMS, Kyoto University}, 21:787--800, 1985.
	\newblock {\href{https://doi.org/10.2977/prims/1195178929}{DOI:
			10.2977/prims/1195178929}}.
	
	\bibitem{Petz-1986}
	D.~Petz.
	\newblock {Quasi-entropies for finite quantum systems}.
	\newblock {\em Reports on Mathematical Physics}, 23(1):57--65, 1986.
	\newblock {\href{https://doi.org/10.1016/0034-4877(86)90067-4}{DOI:
			10.1016/0034-4877(86)90067-4}}.
	
	\bibitem{Petz-1996}
	D.~Petz.
	\newblock {Monotone metrics on matrix spaces}.
	\newblock {\em Linear Algebra and its Applications}, 244:81--96, 1996.
	\newblock {\href{https://doi.org/10.1016/0024-3795(94)00211-8}{DOI:
			10.1016/0024-3795(94)00211-8}}.
	
	\bibitem{Sakai-1956}
	S.~Sakai.
	\newblock {A characterization of $W^{*}$-algebras}.
	\newblock {\em Pacific Journal of Mathematics}, 6(4):763 -- 773, 1956.
	
	\bibitem{takesaki-2002}
	M.~Takesaki.
	\newblock {\em {Theory of Operator Algebra I}}.
	\newblock Springer-Verlag, Berlin, 2002.
	
	\bibitem{Cencov-1982}
	N.~N. \v{C}encov.
	\newblock {\em {Statistical Decision Rules and Optimal Inference}}.
	\newblock American Mathematical Society, Providence, RI, 1982.
	\newblock {\href{https://doi.org/10.1090/mmono/053}{DOI: 10.1090/mmono/053}}.
	
	\bibitem{vomEnde-2020}
	F.~vom Ende.
	\newblock {Strict positivity and D-majorization}.
	\newblock {\em Linear and Multilinear Algebra}, 70(19):4023--4048, 2020.
	\newblock {\href{https://doi.rog/10.1080/03081087.2020.1860887}{DOI:
			10.1080/03081087.2020.1860887},
		\href{https://arxiv.org/abs/2004.05613}{arXiv:2004.05613 [quant-ph]}}.
	
\end{thebibliography}
\end{document}